\documentclass{aastex631}

\usepackage{newtxtext,newtxmath}
\usepackage[T1]{fontenc}
\usepackage{graphicx}	
\usepackage{amsmath}	
\usepackage{amssymb}	
\usepackage{xcolor}
\usepackage{multirow}
\usepackage{subfigure}

\received{April 3, 2023}
\revised{August 27, 2023}
\accepted{October 17, 2023}

\shorttitle{Reconstruction of $H(z)$ using GWs and ML}
\shortauthors{P. Mukherjee, R. Shah, A. Bhaumik, S. Pal}

\begin{document}

\title{Reconstructing the Hubble parameter with future Gravitational Wave missions using Machine Learning}

\correspondingauthor{Purba Mukherjee}
\email{purba16@gmail.com}

\author[0000-0002-2701-5654]{Purba Mukherjee}
\affiliation{Physics and Applied Mathematics Unit \\
Indian Statistical Institute \\
203, B.T. Road, Kolkata 700 108, India}

\author[0000-0001-7682-9219]{Rahul Shah}
\affiliation{Physics and Applied Mathematics Unit \\
Indian Statistical Institute \\
203, B.T. Road, Kolkata 700 108, India}

\author[0000-0002-8421-9397]{Arko Bhaumik}
\affiliation{Physics and Applied Mathematics Unit \\
Indian Statistical Institute \\
203, B.T. Road, Kolkata 700 108, India}

\author[0000-0003-4136-329X]{Supratik Pal}
\affiliation{Physics and Applied Mathematics Unit \\
Indian Statistical Institute \\
203, B.T. Road, Kolkata 700 108, India}
\affiliation{Technology Innovation Hub on Data Science, Big Data Analytics and Data Curation \\
Indian Statistical Institute \\
203, B.T. Road, Kolkata 700 108, India}




\begin{abstract}
We study the prospects of Gaussian processes (GP), a machine learning (ML) algorithm, as a tool to reconstruct the Hubble parameter $H(z)$ with two upcoming gravitational wave missions, namely the evolved Laser Interferometer Space Antenna (eLISA) and the Einstein Telescope (ET). Assuming various background cosmological models, the Hubble parameter has been reconstructed in a non-parametric manner with the help of GP using realistically generated catalogs for each mission. The effects of early-time and late-time priors on the reconstruction of $H(z)$, and hence on the Hubble constant ($H_0$), have also been focused on separately. Our analysis reveals that GP is quite robust in reconstructing the expansion history of the Universe within the observational window of the specific missions under consideration. We further confirm that both eLISA and ET would be able to provide constraints on $H(z)$ and $H_0$ which would be competitive to those inferred from current datasets. In particular, we observe that an eLISA run of $\sim10$-year duration with $\sim80$ detected bright siren events would be able to constrain $H_0$ as good as a $\sim3$-year ET run assuming $\sim1000$ bright siren event detections. Further improvement in precision is expected for longer eLISA mission durations such as a $\sim15$-year time-frame having $\sim120$ events. Lastly, we discuss the possible role of these future gravitational wave missions in addressing the Hubble tension, for each model, on a case-by-case basis.
\end{abstract}



\section{Introduction}\label{sec:introduction}

After the profound success of the LIGO Scientific and Virgo Collaborations in paving the way for gravitational wave (GW) astronomy \citep{ligo_bh,ligo_bns,ligo_multimess,ligo_cat_2017,ligo_cat_2020,ligo_cat_2019}, a number of ground- and space-based GW missions are currently being planned, most of which are expected to be functional within a couple of decades. The space-based Laser Interferometer Space Antenna (LISA) \citep{eLISA,eLISA1,eLISA3} and the ground-based Einstein Telescope (ET) \citep{ET1,ET2} are particularly promising future missions in this regard. Alongside their conventional prospect of contributing to GW astronomy, their roles in cosmology and cosmography are also being investigated to a thorough extent. Designed to explore farther into the redshift space than currently available experiments, these missions are potentially interesting probes of the Hubble parameter as they will be capable of providing an independent measurement of the luminosity distance based on the GW waveforms of detected standard siren events \citep{Schutz1986}. In fact, the pressing need for constraining the Hubble constant ($H_0$) through multi-messenger observations at intermediate redshifts is driven by a fundamental inadequacy of the current cosmological paradigm. 

The concordance $\Lambda$CDM model of cosmology is, of late, facing tensions on multiple fronts when subjected to observational tests. The Hubble tension, being the most severe one among them at $\sim5\sigma$, has lately been a matter of much deliberation. The inconsistency between the measurements of $H_0$ from \textit{Planck} 2018 ($67.36\pm0.54$~km~s\textsuperscript{-1}~Mpc\textsuperscript{-1}) by \citet{Pl18VI} and from direct observations such as SH0ES ($73.30\pm1.04$~km~s\textsuperscript{-1}~Mpc\textsuperscript{-1}) by \citet{Riess_2022} warrants a deep investigation into the systematics and methodologies of the missions, as well as into the assumed cosmological models. On closer inspection, the conflict extends beyond a \textit{Planck}-vs-SH0ES dichotomy and appears to be one between various early time measurements of $H_0$ on one hand and late time measurements on the other. While the \textit{Planck} 2018 best-fit value of $H_0$ sits well with DES clustering and weak lensing data combined with BAO spectroscopic surveys \citep{DESY1BAO}, a number of complementary direct measurement probes besides SH0ES consistently report a higher value \citep{trgb1,trgb2,trgb3,trgb4,mira,mcp2,holicow}. With the systematics being an unlikely culprit \citep{systematics1,systematics2,systematics3}, a plethora of alternatives to the vanilla $\Lambda$CDM model have been proposed to resolve this tension \citep{hubble_hunters_guide,DiValentino_2021,hubble_buyers_guide,To_or_not_to_H0,H0_olympics,Dainotti_2021,Dainotti_2022}. Furthermore, on the basis of some generalized classes of cosmological models, it has also been shown that the Hubble tension might in fact be quite generic to current datasets \citep{Bhattacharyya_2019}. In recent years, these findings have motivated the community to look forward to future GW missions as an alternative probe of $H_0$ at intermediate redshifts, distinct from both early and late-time measurements.

The quantitative scope of constraining $H_0$ using GW data is under active investigation and has shown promise based on existing datasets \citep{LIGO_LSS_1,LIGO_LSS_2,LIGO_H0,LIGOScientific:2017adf,LIGOScientific:2021aug,Fung:2023yyq}, although the limited number of events at present entails wide error bars. Such prospects at present are also, quite expectedly, restricted to fairly low redshifts. For example, \cite{Borhanian:2020vyr} show that the upgraded LIGO and Virgo facilities would be able to constrain the Hubble constant to $\lesssim2\%$ accuracy via a simultaneous measurement of the waveform, and host galaxy redshift from the GW waveform, using dark sirens within $z=0.1$ in five years. On the other hand, \cite{Chen:2017rfc} conclude that a $2\%$ measurement of $H_0$ would be possible in five years, provided the events have reliable electromagnetic (EM) counterparts, or the host galaxy redshifts can be determined using statistical approaches, up to redshifts of $z=0.5$. In order to tackle the degeneracy between source masses and redshifts of binary black hole systems, \cite{Farr:2019twy} propose a novel method of determining the redshifts of such mergers relying on pair-instability supernova processes. They predict that $H(z)$ may be reconstructed at $6.1\%$ in one year, and to $2.9\%$ in five years, at the pivot value of $z=0.8$. For cases where EM counterparts are absent, \cite{Gray:2019ksv} study the effects on inferring redshifts of GW events using incomplete galaxy catalogs. Such lines of inquiry, alongside the inherent limitations arising from sensitivity bounds of the present observatories, naturally lead to the question of how efficient next-generation GW missions might be when it comes to the question of constraining the expansion history of our Universe. Constraints based on future GW data are expected to be much more precise due to significantly higher numbers of detectable events up to higher redshifts, whose prospects have been explored by multiple authors \citep{Messenger:2011gi, Jin:2023sfc,Zhang:2022pnf, Califano:2022syd, Seymour:2022teq, Liu:2022rvk, Ghosh:2022muc, Chen:2022fvf, Cao:2021zpf, Feeney:2020kxk, Mortlock:2018azx, Kyutoku:2016zxn, Jana:2022shb}. However, the predicted results show dependence on the choice of cosmological parameters when real data is absent and forecasts are carried out using mission-specific mock standard siren catalogs \citep{Shah:2023}.

In the absence of a consensus model for cosmic acceleration, there have been attempts to reconstruct the evolutionary history that directly fits with observations without assuming any particular model parametrization (see \citet{Bilicki:2012ub, Ishida_2011, Gomez-Valent:2018gvm, Arjona:2019fwb, Bengaly:2019oxx, Wang:2019vxv, Mukherjee:2020ytg, Mehrabi:2021cob} and references therein). So, a non-parametric reconstruction of the Hubble parameter as a function of redshift using simulated future GW mission data may serve as a powerful predictive tool at this stage, until we have direct data from those missions at hand. Machine learning (ML) algorithms like Gaussian processes (GP) may play an important role for this purpose.

Gaussian processes \citep{rasmussen2006} are a generic supervised learning algorithm designed to solve regression and probabilistic classification problems, depending on whether the output is continuous or discrete. Over the last decade, GP has become particularly popular in cosmology for reconstructing dark energy \citep{Holsclaw_2011, Seikel:2012uu, Zhang:2018gjb} and modified theories of gravity \citep{Zhou_2019,Belgacem_2020,Yang_2021,Bernardo:2021qhu}, exploring the interaction between dark matter and dark energy \citep{Yang:2015tzc,Mukherjee_2021,Cai_2017}, testing the standard concordance model and distance duality relation \citep{Nair:2015jua,Rana:2017sfr,Mukherjee:2021kcu,Mukherjee:2023yxq}, constraining spatial curvature \citep{Yang:2020bpv,Dhawan:2021mel,Mukherjee:2022ujw} and in cosmographic studies \citep{Shafieloo_2012,Mukherjee:2020ytg,Jesus:2022xwb}. For a pedagogical introduction to GP, one can refer to the Gaussian process website\footnote{\url{http://gaussianprocess.org/}}.

In this paper, we propose a novel method to reconstruct the Hubble parameter using Gaussian process regression (GPR) from future gravitational wave mission data. This exercise will further help us investigate the status of the Hubble tension, complementary to the latest available datasets. We take eLISA \citep{eLISA,eLISA1,eLISA3} and ET \citep{ET1,ET2} as two specific examples of next-generation missions to focus on. Taking each mission's specifications into account, we generate mock catalogs of standard sirens for a few representative cosmological models, and employ GPR to reconstruct $H(z)$ and thereby infer $H_0$ directly in a non-parametric manner. Owing to our focus on luminosity distance data, we have specifically chosen models which encapsulate late-time modifications to $\Lambda$CDM. For each of the models, we carry out our analysis on two different sets of catalogs based on two distinct sets of fiducial parameter values, which are obtained by running MCMC for that model on early-time and late-time datasets separately. This disparate treatment is warranted due to the persistent tension between the values of $H_0$ inferred from early and late-time data. In addition to helping us avoid any potential bias which might otherwise result from choosing any particular dataset as sacrosanct in an \textit{a priori} fashion, this methodology allows us to study the dependence of the reconstruction procedure on the chosen priors in a transparent manner. Our study shows that GP is a promising tool which can be used to reconstruct the Hubble parameter $H(z)$, and hence also constrain the Hubble constant $H_0$, using next-generation GW mission data, where 10-years of eLISA results would be at par with 3-years of ET results as far as the precision level is concerned. Besides, the reconstruction results obtained by us also give a concrete estimate of how much the constraints might improve if the eLISA mission runs for a longer duration, say for 15-years.

The paper is organized as follows. In Sec. \ref{sec:reconstruction_steps}, we outline our adopted methodology, present the latest relevant constraints from current datasets that are used as fiducial values, and discuss the procedure for generating the mock standard siren catalogs. In Sec. \ref{sec:gaussian_processes}, we give a brief description of the Gaussian process regression technique and provide the details regarding the implementation. Sec. \ref{sec:analysis_discussions} contains our obtained reconstruction results, followed by a detailed analysis of the same. Finally, we summarize our key findings and make some concluding remarks in Sec. \ref{sec:conclusion}.


\section{Steps towards Reconstruction}\label{sec:reconstruction_steps}

In pursuit of our goal to obtain a non-parametric reconstruction of the Hubble parameter, we implement the steps outlined in the ensuing subsections, taking into consideration the following arguments:

\begin{enumerate}
    \item We generate realistic mock catalogs that incorporate instrumental specifications and account for all possible sources of error for two future GW missions, eLISA and ET separately, to reconstruct the Hubble diagram along with its associated errors. 
    \item In order to simulate a mock catalog, a set of fiducial parameter values must be specified. However, there is no unique choice of fiducial values and this choice may influence the results. Therefore, to select fiducial values that are most realistic, we have taken into consideration the constraints from present datasets on each model.
    \item To ensure that our analysis is unbiased towards any specific realization, as the results may be sensitive to the initial dataset provided, we assemble multiple realizations of these mock catalogs for a particular mission.
    \item The choice of fiducial values may partially draw the mock catalog and hence the reconstructed values towards that particular dataset from which the fiducials were inferred, leading to potential bias in the subsequent analysis. In view of the rising discrepancy between early and late time observations of the Hubble parameter, two different sets of fiducials are considered: one using constraints from early time observations, and the other from late time observations. 
    \item Finally, the results obtained for two different future GW missions, eLISA and ET have been comparatively analyzed. 
\end{enumerate}
By doing so, we believe, the entire analysis would be considered free from any potential bias due to any particular mission or choice of mock catalog, thereby yielding more reliable and robust results for the GW missions under consideration.

\setcounter{table}{0}
\begin{table}[!ht]
\renewcommand{\thetable}{\arabic{table}}
\centering
\caption{Brief summary of the models/parametrizations under consideration. Here "Additional parameters" means parameters on top of 6.}\label{tab:models}
\resizebox{\textwidth}{!}{\renewcommand{\arraystretch}{1.2} \setlength{\tabcolsep}{15 pt}
\begin{tabular}{p{6cm}|c|p{7cm}} 
\hline
\textbf{Models} & \textbf{Additional parameters} & \textbf{Motivation} \\
\hline \hline
$\Lambda$ Cold Dark Matter ($\Lambda$CDM) &  $-$  & Baseline (6-parameter) \\
\hline
Chevallier-Polarski-Linder (CPL) \citep{cpl1, cpl2} &  $w_0$, $w_a$  &  Simplest dynamical DE parametrization with redshift-varying EoS to resolve the coincidence problem: $w(z)=w_0+w_a\frac{z}{1+z}$.  \\
\hline
Jassal-Bagla-Padmanabhan (JBP) \citep{jbp}  & $w_0$, $w_a$ & Phenomenologically chosen redshift-dependent DE EoS: $w(z)=w_0+w_a\frac{z}{\left(1+z\right)^2}$. \\
\hline
Phenomenologically Emergent Dark Energy (PEDE) \citep{Li_2019} & $-$ &  Redshift-dependent DE density parameter to resolve the $H_0$ tension: $\Omega(z)=(1-\Omega_{m0}) [1-\tanh(\log_{10}(1 + z))]$.  \\
\hline
Vacuum Metamorphosis (VM) \citep{vm1, vm2, vm3}  &  $M$  related to $\Omega_{m0}$ &  Motivated by non-perturbative effects of quantum gravity where a gravitational phase transition occurs at some critical redshift $z_c=-1+\frac{3\Omega_{m0}}{4(1-M)}$. \\
\hline
Elaborated Vacuum Metamorphosis (VM-VEV) \citep{VM_Di_Valentino_2020} &  $M$ independent of $\Omega_{m0}$ &  Extension of original VM model allowing a non-vanishing DE component at $z>z_c$, i.e., before the gravitational phase transition. \\ 
\hline
\end{tabular}
}
\end{table}    

\setcounter{table}{1}
\begin{table}[!ht]
\renewcommand{\thetable}{\arabic{table}}
\centering
\caption{Table showing a summary of the observational datasets utilized to obtain the latest constraints on the different models/parametrizations, and the gravitational wave missions considered for the analysis.}\label{tab:study_cases}
\resizebox{\textwidth}{!}{\renewcommand{\arraystretch}{2.1} \setlength{\tabcolsep}{5 pt} 
\begin{tabular}{>{\centering\arraybackslash}p{3cm} | >{\centering\arraybackslash}p{9.1cm} | >{\centering\arraybackslash}p{5.0cm} >{\centering\arraybackslash}p{5.0cm}}
\hline
\multicolumn{2}{c|}{\textbf{Current Cosmological Surveys}} &  \multicolumn{2}{c}{\textbf{Upcoming GW Missions}}  \\ 
\cline{1-4}
\multirow{1.9}{1.8cm}{\textbf{Compilation}} & \multirow{1.9}{3.15cm}{\textbf{Observational Datasets}} &  \textbf{evolved Laser Interferometer Space Antenna} (eLISA) \citep{eLISA} & \textbf{Einstein Telescope} (ET) \citep{ET1, ET2} \\
\hline \hline 
\multirow{2.5}{0.5cm}{CSB} & \multirow{2.8}{9.cm}{\textbf{Cosmic Microwave Background} : \\
\textit{Planck} 2018 \textit{TTTEEE+low l+low E+lensing} \citep{Pl18V, Pl18VI}, \\ \textbf{Baryon Acoustic Oscillations} (BAO): \\
6dFGS \citep{6dF}, SDSS MGS \citep{MGS}, BOSS DR12 \citep{BOSS}, \\
\textbf{Type Ia supernovae}: Pantheon \citep{Pantheon}} \\ &  & \multirow{1.}{2cm}{\centering eLISA Set-I} &  \multirow{1.}{2cm}{ET Set-I}\\  
\hline
\multirow{2.5}{0.5cm}{RSH} & \multirow{2.8}{9.cm}{\textbf{S}upernovae, \textbf{H0}, for the \textbf{E}quation of \textbf{S}tate of \textbf{D}ark energy (SH0ES) \citep{Riess_2022}, \\ 
\textbf{Cosmic Chronometers} (CC): \citep{stern2010, moresco2012, zhang2014, moresco2015b, moresco2016, ratsim2017, borghi2022}, \\ \textbf{Type Ia supernovae}: Pantheon \citep{Pantheon}} \\ &  & \multirow{1.}{1.8cm}{eLISA Set-II} & \multirow{1.}{2cm}{ET Set-II} \\
\hline
\end{tabular}
}
\end{table}    

\subsection{Fiducial values from current constraints}\label{sec:fiducials}

We consider six representative cosmological models for the purposes of this study, as given in Table \ref{tab:models}. By identifying the vanilla $\Lambda$CDM model as the baseline $6$-parameter model, these can be classified into $6$, $(6+1)$, and $(6+2)$ -parameter scenarios. Besides $\Lambda$CDM, five other models/parametrizations have been chosen based primarily on their potential in addressing the Hubble tension in view of present data \citep{DiValentino_2021}. A brief discussion on each model and their tension-resolving potential in light of future missions have been explored in \citet{Shah:2023}. The present article focuses on a nearly unbiased investigation of GP as a tool to reconstruct $H(z)$ from next-generation GW data. However, the value of $H_0$ is obtained from the analysis as a natural consequence.

Given the discrepancies in the constraints on the parameters of the concordance model when subjected to late-time experiments versus early-time experiments, we choose not to assume either set of constraints to be the absolutely correct one. We rather split the currently available experimental data into two compilations - namely, CSB, from the early-time probes and RSH, from observations in the late Universe. Table \ref{tab:study_cases} gives a summary of the existing datasets and future GW missions under consideration. Using these two combinations of datasets, we first bring all six cosmological models under consideration to an equal status when it comes to their parameter constraints from the latest observational probes.

The MCMC constraints for each model corresponding to CSB data have already been investigated partially by the present authors \citep{Shah:2023} and partially by others \citep{VM_Di_Valentino_2020}. We run the Boltzmann code CLASS \citep{CLASS1,CLASS2} and the Markov chain Monte Carlo (MCMC) parameter estimation code MontePython \citep{MontePython1,MontePython2} to obtain the constraints for the same models with RSH data. Table \ref{constraints} gives a summary of the above parameter constraints. We consider the following uniform priors for each cosmological model parameter when running MCMC: $H_0\in[40,110]$, $\Omega_{m0}\in[0.01,0.99]$, $w_0\in[-2,1]$, $w_a\in[-3,3]$ and $M\in[0.5,1]$.

We consider the two sets individually in our analysis to focus on both ends of the spectrum when it comes to the tension between early-time and late-time inferences and implement the constraints thus obtained as fiducial values for the mock standard siren catalogs which we generate at the next step.

\setcounter{table}{2}
\begin{table}[!t]
	\renewcommand{\thetable}{\arabic{table}}
	\centering
    \caption{Latest constraints on the cosmological model parameters using combined CMB+Pantheon+BAO (CSB) and SH0ES+Pantheon+CC (RSH) observational data. [Constraints marked with $\dagger$ have been quoted from \citet{Shah:2023}, and * from \citet{VM_Di_Valentino_2020} respectively].}\label{constraints}
    \resizebox{1.0\textwidth}{!}{\renewcommand{\arraystretch}{2} \setlength{\tabcolsep}{11 pt} 
    \begin{tabular}{c c c c c c c c}
        \hline
        \textbf{Compilation} & \textbf{Parameters} & $\boldsymbol{\Lambda}$\textbf{CDM}$^\dagger$ & \textbf{CPL}$^\dagger$ & \textbf{JBP}$^\dagger$ & \textbf{PEDE}$^\dagger$ & \textbf{VM}* & \textbf{VM-VEV}* \\ \hline
        \hline
     &   $H_0$ & $67.72_{-0.41}^{+0.42}$ & $68.34_{-0.88}^{+0.83}$ & $68.32_{-0.82}^{+0.78}$ & $71.24_{-0.48}^{+0.49}$ & $74.21_{-0.66}^{+0.66}$ & $73.26_{-0.32}^{+0.32}$  \\
     &   $\Omega_{m0}$ & $0.3102_{-0.0057}^{+0.0054}$ & $0.3064_{-0.0081}^{+0.0079}$ & $0.3062_{-0.0078}^{+0.0075}$ & $0.2855_{-0.0056}^{+0.0051}$ & $0.2593_{-0.0046}^{+0.0046}$ & $0.2695_{-0.0041}^{+0.0041}$  \\
CSB  &   $w_0$ & - & $-0.9571_{-0.082}^{+0.078}$ & $-0.9705_{-0.12}^{+0.12}$ & - & - & -  \\
     &   $w_a$ & - & $-0.2904_{-0.28}^{+0.33}$ & $-0.3648_{-0.78}^{+0.74}$ & - & - & -  \\
     &   $M$ & - & - & - & - & $0.9277_{-0.0044}^{+0.0044}$ & $0.8929_{-0.0016}^{+0.0010}$  \\
        \hline
        \textbf{Compilation} &  \textbf{Parameters} & $\boldsymbol{\Lambda}$\textbf{CDM} & \textbf{CPL} & \textbf{JBP} & \textbf{PEDE} & \textbf{VM} & \textbf{VM-VEV} \\ \hline
        \hline
     &   $H_0$ & $72.84_{-1.00}^{+1.00}$ & $72.80_{-1.00}^{+1.00}$ & $72.84_{-1.00}^{+1.05}$ &  $72.71_{-0.99}^{+1.00}$  &  $72.42^{+0.96}_{-0.96}$ & $72.80^{+0.97}_{-0.97}$  \\
     &   $\Omega_{m0}$ & $0.2909_{-0.019}^{+0.019}$ & $0.2820_{-0.0455}^{+0.0470}$ &  $0.2830_{-0.0470}^{+0.0530}$ &  $0.3334_{-0.018}^{+0.018}$  & $0.3533^{+0.0214}_{-0.0213}$ & $0.3007^{+0.0235}_{-0.0221}$  \\
RSH  &   $w_0$ & - & $-1.008_{-0.129}^{+0.149}$ & $-1.004_{-0.149}^{+0.160}$ & - & - & -  \\
     &   $w_a$ & - & $0.0468_{-0.42}^{+0.38}$ & $0.0431_{-0.57}^{+0.51}$ & - & - & -  \\
     &   $M$ & - & - & - & - & $0.8767^{+0.0103}_{-0.0105}$ & $0.8102^{+0.0213}_{-0.0216}$  \\
        \hline
    \end{tabular}
    }
\end{table}

\subsection{Generation of mock GW catalogs}\label{sec:mocks}

ET, a proposed third-generation ground-based European GW detector, is an evolution of 2G detectors such as Advanced LIGO, Advanced Virgo, and KAGRA \citep{ET1}. As demonstrated by \citeauthor{Belgacem:2018lbp}, ET is expected to observe $\sim10^3$ binary neutron star (BNS) merger events with electromagnetic counterparts from associated short-hard $\gamma$-ray bursts over an observation period of 3-years. Synergies with instruments like the Transient High-Energy Sky and Early Universe Surveyor (THESEUS) \citep{THESEUS:2017qvx}, the Extremely Large Telescope (ELT) \citep{eltbook}, the Advanced Telescope for High Energy Astrophysics (ATHENA+) \citep{Piro:2021oaa}, and the Daksha satellites \citep{Bhalerao:2022edb} can help in rapid detection of the EM counterpart transients and subsequent determination of the redshifts of the host galaxies. Alternative methods include employing advanced statistical techniques to estimate the redshifts of associated galaxies \citep{ETMukherjee:2018ebj}, or cross-correlating GW catalogs with large-scale structure observations \citep{ETScelfo:2018sny}. ET is expected to observe events from a minimum of $z=0.07$ to a maximum of $z\sim2$ with SNR $>8$. The lower cutoff is maintained to exclude sources which require modelization of the local Hubble flow before including them in the analysis \citep{ET3}. To generate our event catalog, we use the standard expression for the number density of the observed events within some redshift interval \citep{ET2}, which is a function of the coalescence rate at a given redshift. This rate is obtained via a fit to the observationally determined star formation history \citep{Schneider:2000sg}. We generate $d_L$ samples according to these distributions. To get an estimate of the uncertainties on $d_L$, we add in quadrature the instrumental error \citep{ET3} and the error due to lensing \citep{ET2}.

The space-based eLISA mission is expected to revolutionize our understanding of cosmic history by probing massive black hole binaries (MBHBs). High SNR detection of MBHB coalescence will yield accurate direct measurements of their luminosity distance. Additionally, simultaneous detection of EM counterparts would be possible for events with SNR $>8$ and a sufficiently small sky localization error \citep{eLISA3,Dotti_2012,Antonini:2015sza}. Such EM counterparts may be detected either by the Large Synoptic Survey Telescope (LSST) \citep{lsst} in the optical range, or through the joint efforts of the Square Kilometre Array (SKA) \citep{ska} and ELT in the radio frequency range. The former would be possible for quasar-like luminosity flares which might be triggered during mergers due to the presence of sufficient amounts of surrounding gas, whereas the latter would be the way to go for radio flares and jets due to mergers occurring in the presence of sufficiently strong galactic magnetic fields. \citeauthor{eLISA3} have shown that the eLISA mission (in the L6A2M5N2 configuration \citealt{eLISA}) is expected to observe $\sim25$ MBHB coalescence events for 'No Delay' \citep{heavyseed1,heavyseed3,eLISA1} sources over a mission duration of 3-years up to redshifts $z\sim7-8$. A comparative study of the number of detectable sources with redshift measurable counterparts to populate the $d_L-z$ diagram is given in Tables 9 and 10 of \citet{eLISA3}. To generate mock catalogs, we draw plausible redshift values for MBHB mergers from an interpolated $\beta$ distribution, based on the data presented in Fig. 1 of \citet{eLISA3}. The lower bound is introduced at $z<0.1$, to match our expectations of not finding MBHBs at very low redshift \citep{Speri:2020hwc}. To estimate a realistic 1$\sigma$ error on $d_L$ to each MBHB merger event, we combine the uncertainties contributed by lensing \citep{Tamanini:2016uin, Hirata_2010, Shapiro_2010}, peculiar velocity corrections \citep{Tamanini:2016uin, Kocsis:2005vv}, instrumental precision \citep{Marsat:2020rtl, Speri:2020hwc}, and redshift error associated with photometric measurements \citep{Dahlen:2013fea, Ilbert:2013bf}.

Keeping these assumptions and arguments in mind, we generate mock catalogs so as to give a reliable number of data points for each mission. In order to get rid of any potential bias from a particular mock catalog, we generate 500 mock catalogs of observable bright siren events for each case in our study (see Table \ref{tab:study_cases}). We also make some modifications to the methodology outlined in \citet{Ferreira} to make the catalogs more realistic as mentioned in \citet{Shah:2023}. Here is a brief outline of our adopted methodology:

\begin{itemize}
    \item Obtain the set of redshifts of bright siren events for both eLISA and ET by sampling from the theoretical redshift distribution based on the specifications of each mission.
    \item Choosing a particular cosmological model, calculate the theoretical luminosity distance $d_L^{\{th\}}(z)$ at the sampled redshifts.
    \item Estimate the total error $\Delta\nobreak{}d_L(z)$ by taking into account the different sources of error which affect the measurement of the luminosity distance for each mission.
    \item Obtain the simulated $d_L(z)$ by randomly sampling from a normal distribution $\mathcal{N}(d_L^{\{th\}}(z),\Delta\nobreak{}d_L(z))$.
\end{itemize}

This gives us a set of event redshifts $\{z\}$, the corresponding luminosity distances $\{d_L(z)\}$, and the observational errors in determining the latter $\{\Delta\nobreak{}d_L(z)\}$. Such a set constitutes an individual catalog corresponding to each combination of mission, cosmological model and observational prior.


\section{Gaussian Process Reconstruction}\label{sec:gaussian_processes}

Let us begin with a brief description of the ML tool, namely the Gaussian process \citep{rasmussen2006,Seikel:2012uu,Shafieloo_2012}, that we are going to use in our analysis. A Gaussian process involves an indexed collection of random variables having a multivariate normal (MVN) distribution, used to infer a distribution over functions directly within a continuous domain. For a given set of observational data, one can use GP to reconstruct the most probable underlying continuous function describing that data and the associated confidence levels, without limiting to any particular parametrization ansatz. Thus, GP serves as a powerful, non-parametric tool for statistical modeling.

For undertaking GPR, our assumption is that the observational data with Gaussian noise, as well as the predicted function, jointly underlie an MVN distribution, described solely by a mean vector and a covariance matrix. Thus, given a mock $d_L$ vs $z$ catalog, we can employ GP to reconstruct $d_L(z)$ and $\sigma_{d_L}(z)$ directly from data, without assuming any model parametrization. The reconstructed function at some redshift $z$ is characterized by a mean $\mu(z)=\mathbf{E}[d_L(z)]$ and covariance function (also called ``kernel") $\text{cov}(d_L(z),d_L(\tilde{z}))=\kappa(z,\tilde{z})$, which correlates values of the function between two redshifts, $z$ and $\tilde{z}$, respectively. As prior information for the predictions, we assume a zero mean function to ensure a cosmological model-independent analysis. Hence, the choice of kernel determines almost all the generalization properties of our GPR.

A wide range of possible covariance functions is available in the literature \citep{rasmussen2006, Seikel:2012uu}. The most general and standard choice is the squared exponential covariance,
\begin{equation}
    \kappa(z,\tilde{z})=\sigma_f^2\exp\left[-\frac{\left(z-\tilde{z}\right)^2}{2l^2}\right],
\end{equation}
where, $l$ and $\sigma_f$ are the characteristic length scale and noise variance, known as the \textit{hyperparameters} of our GP model. Note $l$ roughly corresponds to the distance one needs to move along $z$ before $d_L$ significantly changes, whereas $\sigma_f$ describes typical changes in the value of $d_L$. 

Another possible choice is the Mat\'{e}rn covariance, with $\nu\equiv\nobreak{}p+\frac{1}{2}$, given by,
\begin{equation}\label{ch1:matern}
\kappa_{\nu=p+\frac{1}{2}}(z,\tilde{z})=\sigma_f^2\exp\left(\frac{-\sqrt{2p+1}}{l}\vert\nobreak{}z-\tilde{z}\vert\right)\frac{p!}{(2p)!}{{\sum}}_{i=0}^{p}~\frac{(p+i)!}{i!(p-i)!}\left(\frac{2\sqrt{2p+1}}{l}\vert\nobreak{}z-\tilde{z}\vert\right)^{p-i}, 
\end{equation} 
where $p$ denotes the order. 

The squared exponential kernel is indefinitely differentiable and for a reconstruction involving an $n$th order derivative, the Mat\'{e}rn $\nu$ covariance works well if $\nu>n$. Note that the reconstruction kernel plays a significant role and the final results are sensitive to this choice. Imposing greater differentiability, hence a greater degree of smoothness, leads to strong correlations in the reconstructed functions and their derivatives, resulting in tighter bounds on the uncertainties. As the present work requires the use of second-order kernel derivatives, in this work, we choose the Mat\'{e}rn $\nu=9/2$ covariance, 
\begin{equation}
\kappa(z,\tilde{z})=\sigma_f^2e^{-\frac{3\vert\nobreak{}z-\tilde{z}\vert}{l}}\left[1+\frac{3\vert\nobreak{}z-\tilde{z}\vert}{l}+\frac{27(z-\tilde{z})^2}{7l^2}+\frac{18\vert\nobreak{}z-\tilde{z}\vert^3}{7l^3}+\frac{27\left(z-\tilde{z}\right)^4}{35l^4}\right],
\end{equation}
as suggested in \cite{Seikel:2013fda}. Moreover, it has been shown in \cite{OColgain:2021pyh} that the reconstructed errors decrease as $\nu \rightarrow \infty$ (squared exponential kernel), but the difference between $\nu=5/2$ and $\infty$ is not highly striking. With the marginalized covariance function, the data can be extended to the redshift range $z$, specific to a particular mission, for obtaining a smooth continuous function for $d_L$ with respective uncertainties $\sigma_{d_L}$. 

As a demonstrative example, consider a single eLISA mock catalog generated assuming the baseline $\Lambda$CDM model. We have used a \texttt{python}-implementation for GP, being simple linear algebra. In the next section, we will extensively make use of this algorithm for other catalogs. Given a set of training inputs $Z=\left\lbrace{z_i}\right\rbrace$ from the $d_L(Z)$ vs $Z$ GW mock catalog (Sec. \ref{sec:mocks}) with noise $\sigma_{d_L}(Z)$, we use GPR to make predictions of $d_L(Z^\star)$ on a test set of redshifts $Z^\star=\left\lbrace{z^\star_i}\right\rbrace$. The prior covariance matrix between the training points $\kappa(Z,Z)$ is defined as $\left[\kappa(Z,Z)\right]_{ij}$=$\kappa(z_i,z_j)$. Similarly, we also obtain the prior covariances between the training vs test and test vs test redshifts as $\kappa(Z,Z^\star)$ and $\kappa(Z^\star,Z^\star)$ respectively. So, the joint prior distribution of the training data and test predictions of $d_L$ is given by $\sim\mathcal{N}(0,\Sigma)$, where,
\begin{equation}
\Sigma=\begin{bmatrix}\kappa(Z,Z)+\mathcal{C}&\kappa(Z,Z^\star)\\\kappa(Z^\star,Z)&\kappa(Z^\star,Z^\star)\end{bmatrix}, 
\end{equation} 
and $\mathcal{C}$ is the covariance matrix of the mock data.

\begin{figure}[!t]
\gridline{\fig{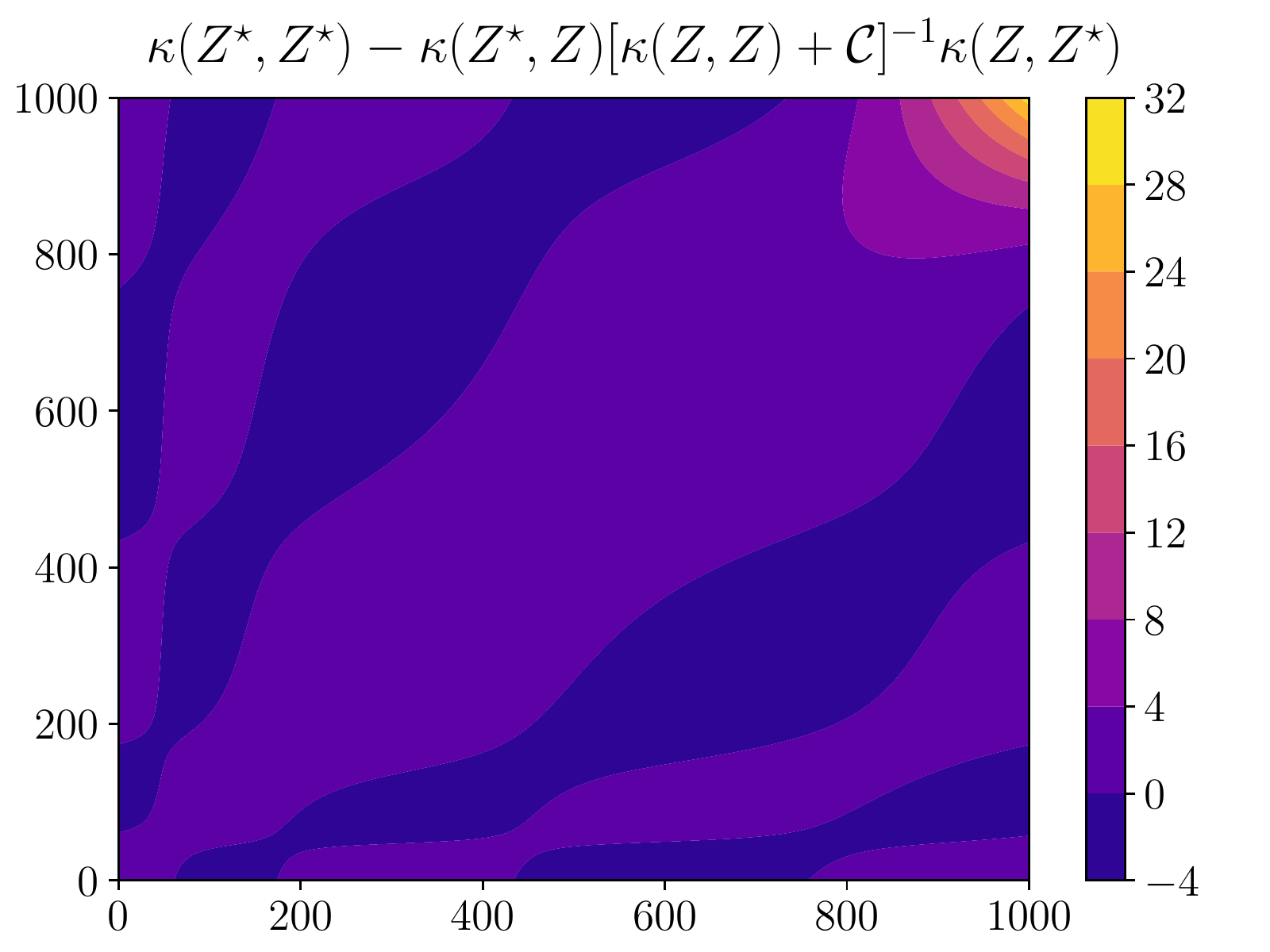}{0.3\textwidth}{(a)}
          \fig{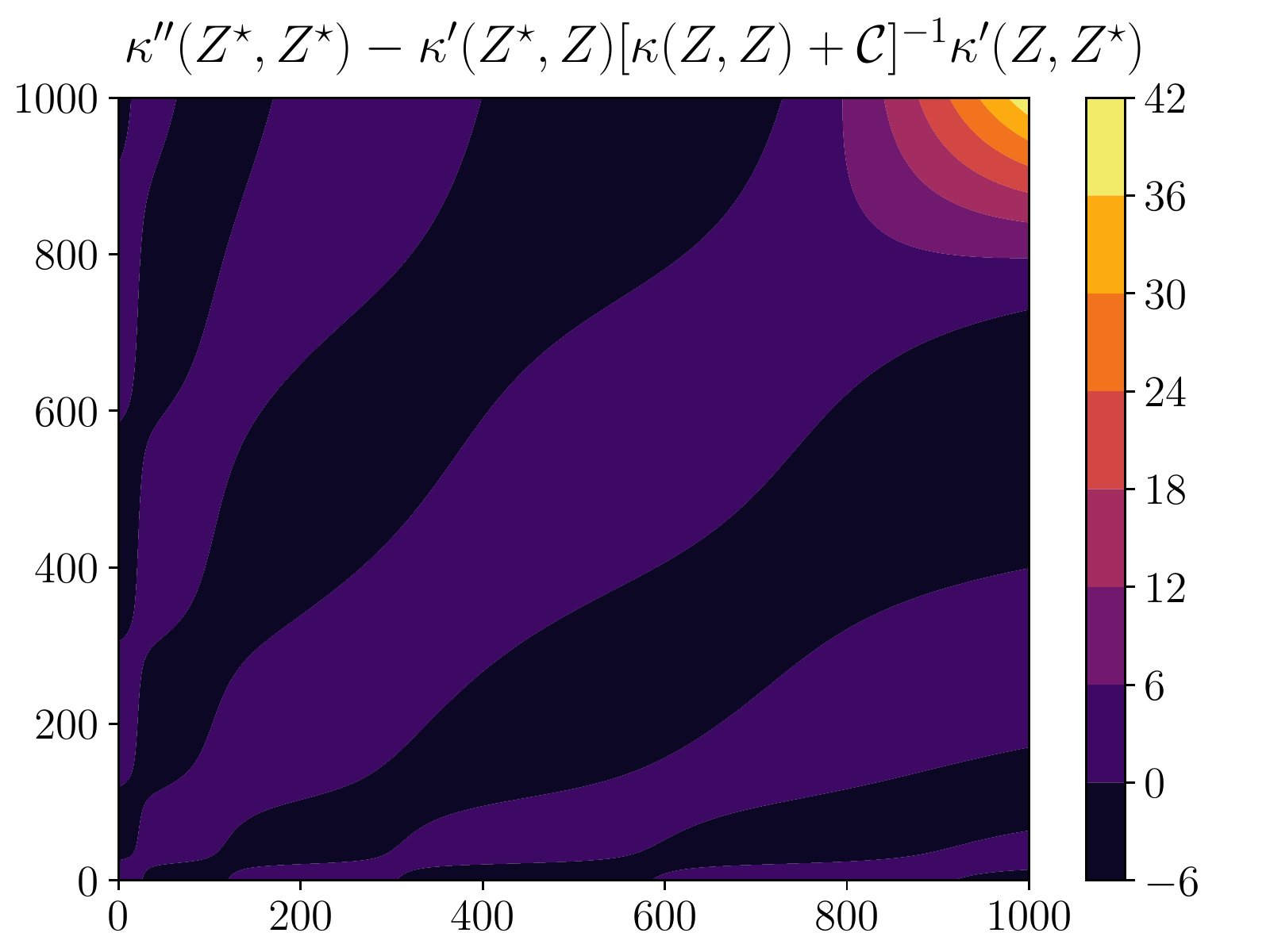}{0.3\textwidth}{(b)}
          \fig{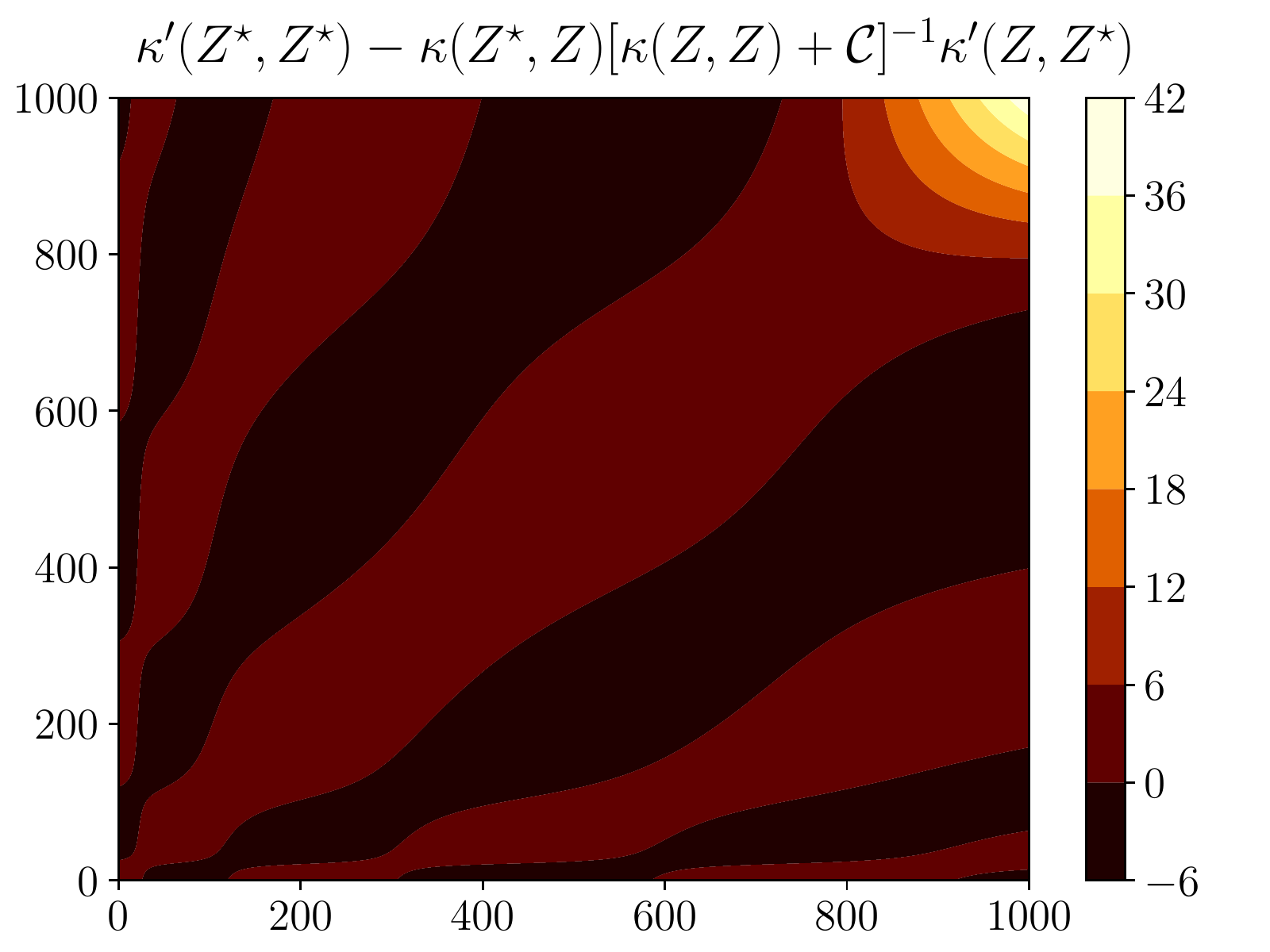}{0.3\textwidth}{(c)}} \vskip -0.3cm
\caption{Plot for the posterior covariance matrices (a) $\text{cov}[d_L(Z^\star),d_L(Z^\star)]$, (b) $\text{cov}[d_L'(Z^\star),d_L'(Z^\star)]$, (c) $\text{cov}[d_L(Z^\star),d_L'(Z^\star)]$, defined in Eq. \eqref{eq:posterior_cov_dldl}, \eqref{eq:posterior_cov_dlpdlp} and \eqref{eq:posterior_cov_dldlp} respectively.}\label{fig:pred_cov}
\end{figure}

The predictive distribution is given by $\sim\mathcal{N}\left(d_L(Z^\star),\text{cov}[d_L(Z^\star),d_L(Z^\star)]\right)$, where the predicted mean vector, $d_L(Z^\star)$ and the covariance matrix cov[$d_L(Z^\star),d_L(Z^\star)$] are,
\begin{equation}\label{eq:posterior_dl}
d_L(Z^\star)=\kappa(Z^\star,Z)\left[\kappa(Z,Z)+\mathcal{C}\right]^{-1}d_L(Z)\:,
\end{equation} 
and,
\begin{equation}\label{eq:posterior_cov_dldl}
\text{cov}[d_L(Z^\star),d_L(Z^\star)]=\kappa(Z^\star,Z^\star)-\kappa(Z^\star,Z)[\kappa(Z,Z)+\mathcal{C}]^{-1}\kappa(Z,Z^\star).
\end{equation} 

Again, GP can also be used to reconstruct the derivatives of $d_L$, assuming these derivatives also follow a joint MVN with the observed data. The mean vector and the covariance matrix corresponding to the first derivative ${d_L}^{\prime}$ are given by,
\begin{equation}\label{eq:posterior_dlp}
d_L^\prime(Z^\star)=\kappa'(Z^\star,Z)\left[\kappa(Z,Z)+\mathcal{C}\right]^{-1}d_L(Z)\:,
\end{equation} 
and,
\begin{equation}\label{eq:posterior_cov_dlpdlp}
\text{cov}[d_L^\prime(Z^\star),d_L^\prime(Z^\star)]=\kappa''(Z^\star,Z^\star)-\kappa'(Z^\star,Z)[\kappa(Z,Z)+\mathcal{C}]^{-1}\kappa'(Z, Z^\star).
\end{equation}

Furthermore, the covariance between the reconstructed functions ${d_L}(Z^\star)$ and ${d_L}^{\prime}(Z^\star)$ can be obtained as,
\begin{equation}\label{eq:posterior_cov_dldlp}
\text{cov}[d_L(Z^\star),d_L^\prime(Z^\star)]=\kappa'(Z^\star,Z^\star)-\kappa(Z^\star,Z)[\kappa(Z,Z)+\mathcal{C}]^{-1}\kappa'(Z, Z^\star).
\end{equation}

Here, a prime denotes a derivative of the prior covariance function with respect to redshift,
\begin{align}\label{eq:deriv_K_priors}
\kappa'(Z,Z^\star)=\frac{\partial}{\partial Z}\kappa(Z,Z^\star)\,,\,\,\,\kappa'(Z^\star,Z^\star)&=\frac{\partial}{\partial Z^\star}\kappa(Z^\star,Z^\star)\,,\,\,\,\kappa''(Z^\star,Z^\star)=\frac{\partial^2}{\partial{Z^\star}\partial{Z^\star}}\kappa(Z^\star,Z^\star).\nonumber
\end{align}

In what follows we make use of the above-mentioned algorithm for non-parametric reconstruction of the Hubble parameter. We also need to know the hyperparameters $\sigma_f$ and $l$, which are obtained by optimizing the log-marginal likelihood, which in this case is given by,
\begin{equation}\label{eq:lnlike}
\ln\mathcal{L}=-\frac{1}{2}d_L(Z)^{\text{T}}[\kappa(Z,Z)+\mathcal{C}]^{-1}d_L(Z)-\frac{1}{2}\ln\vert\kappa(Z,Z)+\mathcal{C}\vert-\frac{n}{2}\ln 2\pi.
\end{equation}

\begin{figure}[!t]
\gridline{\fig{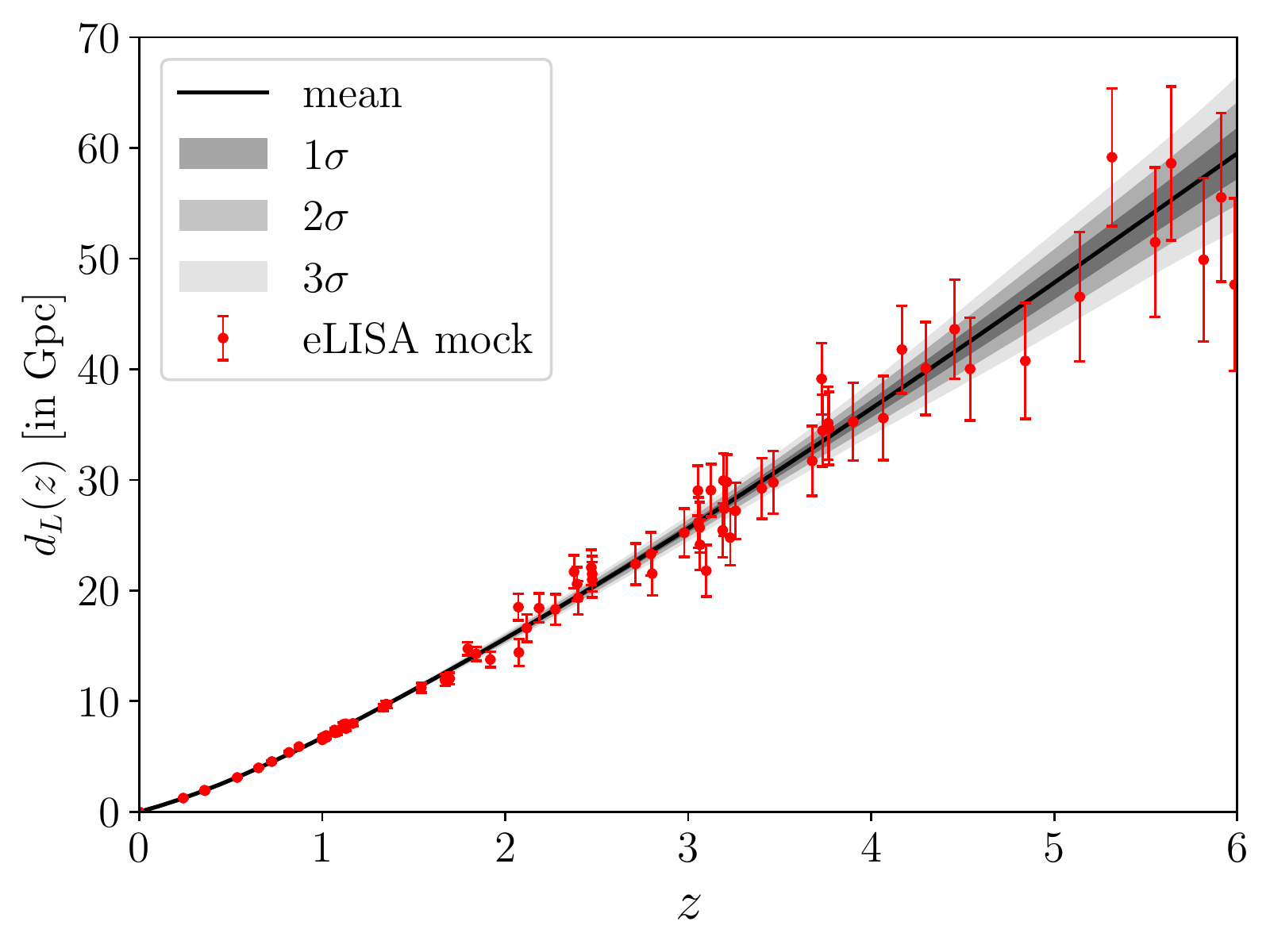}{0.3\textwidth}{}
          \fig{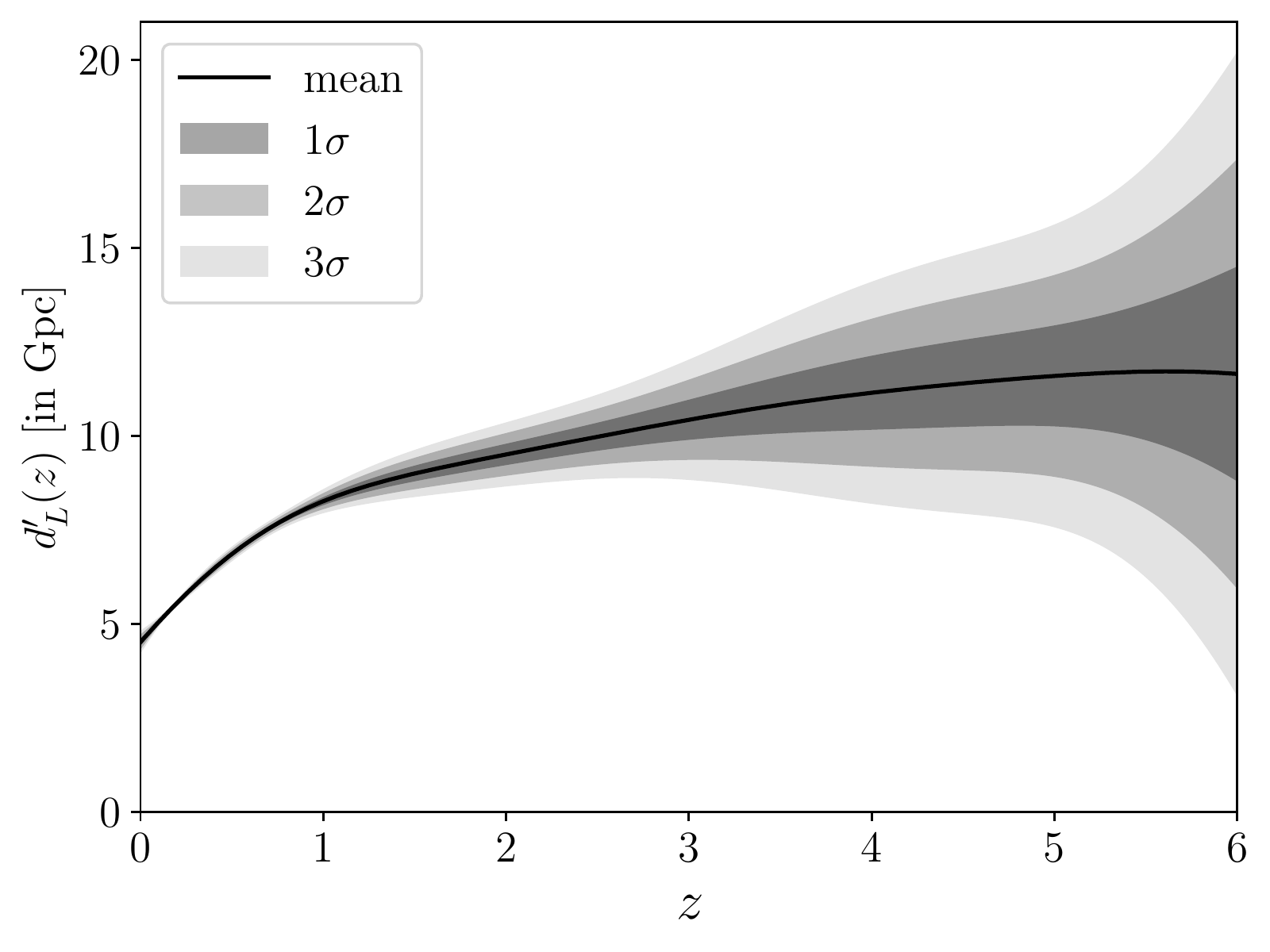}{0.3\textwidth}{}
          \fig{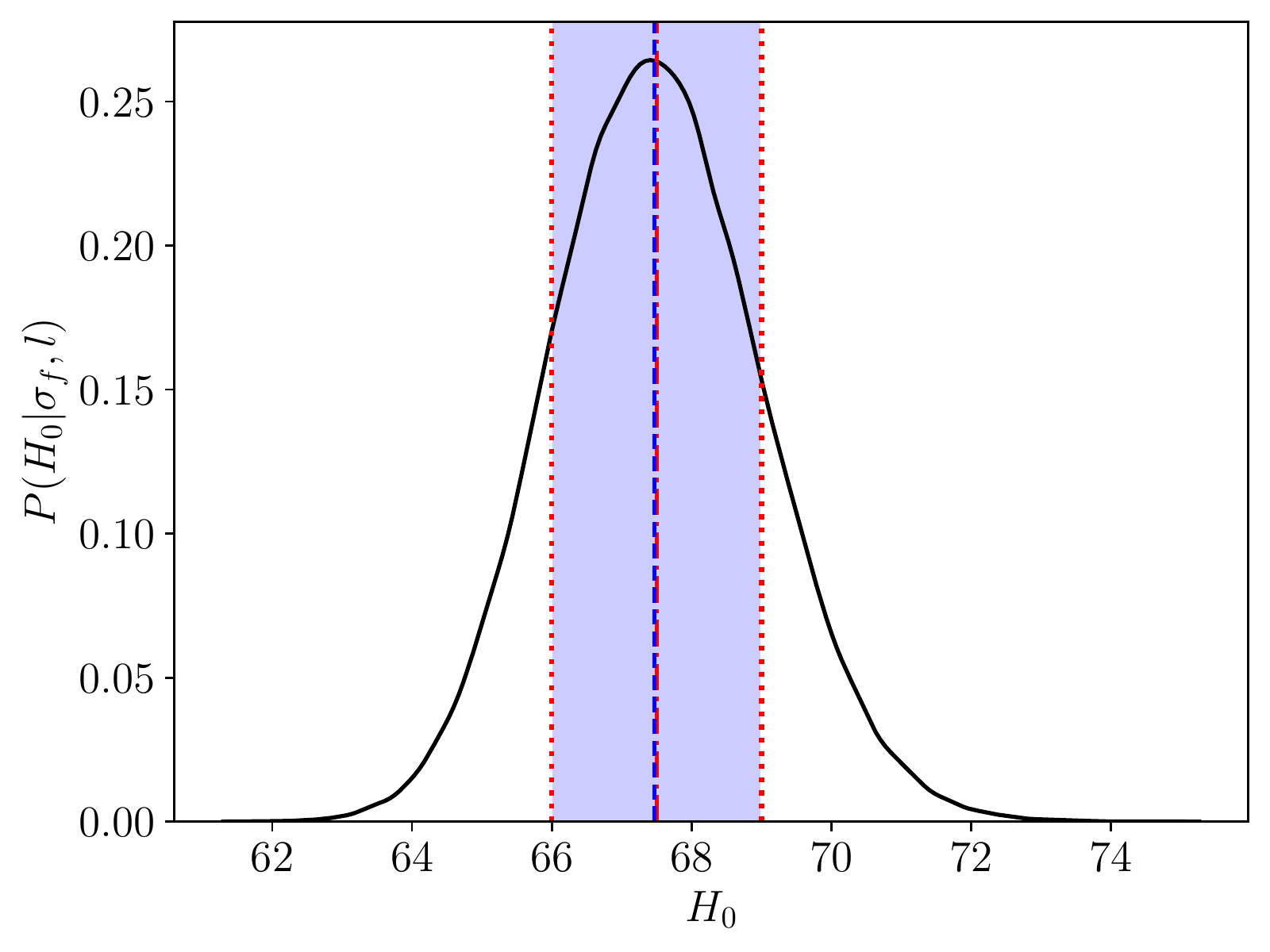}{0.3\textwidth}{}}\vskip-0.7cm
\caption{Plots showing the reconstructed functions $d_L$ (left panel) and $d_L'$ (middle panel) in units of Gpc with respect to the test redshifts. The right panel shows the reconstructed posterior distribution of $H_0$, $P(H_0\vert\sigma_f,l)$, for the best-fit values of the hyperparameters.}\label{fig:Dl_plot}
\end{figure}

For a complete Bayesian analysis, we marginalize the log-likelihood (Eq. \eqref{eq:lnlike}) over the hyperparameters instead of optimizing them.  So, the posterior predictive distribution is,
\begin{equation}
p(\sigma_f,l\vert\nobreak{}Z,d_L(Z),\mathcal{C})\propto\nobreak{}p(\sigma_f,l)~\mathcal{L}~,
\end{equation} 
where $p(\sigma_f,l)$ is the `prior' function that we take as uniform. We adopted a \texttt{python} implementation of the ensemble sampler for MCMC, \texttt{emcee}, introduced by \citep{Foreman_Mackey_2013}.

Using the best-fit samples from the posterior distribution of hyperparameters, we make predictions for the functions $d_L$ and $d_L'$ at the test redshifts $Z^\star\equiv\left\lbrace\nobreak{}z\right\rbrace$ with the help of Eq. \eqref{eq:posterior_dl} and \eqref{eq:posterior_dlp} assuming $1000$ reconstruction bins along redshift range $0<z<z_{\text{max}}$. We plot posterior covariance matrices $\text{cov}[d_L,d_L], \text{cov}[d_L^\prime,d_L^\prime]$ and $\text{cov}[d_L,d_L^\prime]$ in Fig. \ref{fig:pred_cov}. The reconstructed functions ${d_L}(z)$ and ${d_L}^{\prime}(z)$ also shown in Fig. \ref{fig:Dl_plot}. 

Finally, with these reconstructed $d_L(z)$ and ${d_L}^{\prime}(z)$, we derive the evolution of the Hubble parameter as,
\begin{equation}
    H(z)=\frac{c(1+z)^2}{{d_L}^{\prime}(1+z)-d_L},
\end{equation}
and infer $H_0\equiv H(z=0)$ directly in a non-parametric way. The uncertainties associated with the reconstructed $H_0$ are estimated via the Monte Carlo error propagation technique. We show the posterior distribution of the Hubble parameter at present epoch $H_0$ for a given sample of hyperparameters, i.e., $P(H_0\vert\sigma_f,l)$ with respect to $H_0$ in the right panel of Fig. \ref{fig:Dl_plot}. The best-fit $H_0$ is shown with a blue dashed line. The shaded region in blue is the standard deviation. The mean $H_0$ along with the 1$\sigma$ uncertainties are plotted in red with dash-dot and dotted lines respectively. To validate our results, we compared the plots shown in Fig. \ref{fig:Dl_plot} to those obtained from using the \texttt{GaPP} \citep{Seikel:2012uu} code.

Now, for each of the representative models mentioned in Table \ref{tab:study_cases}, we employ GP to directly reconstruct $d_L(z)$, ${d_L}^\prime(z)$ and thereby obtain the evolution of $H(z)$ as a function of redshift using the respective 500 mock catalogs, generated in Sec. \ref{sec:mocks}, separately for both early-time (CSB) and late-time (RSH) fiducial values. Henceforth, we shall refer to them as `Set-I' and `Set-II' respectively. For this exercise, we take into account two GW missions under consideration, namely, eLISA and ET, separately, and finally compute the averaged reconstructed $H(z)$ functions from the respective mock catalog compilations.

\begin{figure}[!t]
\includegraphics[width=\textwidth]{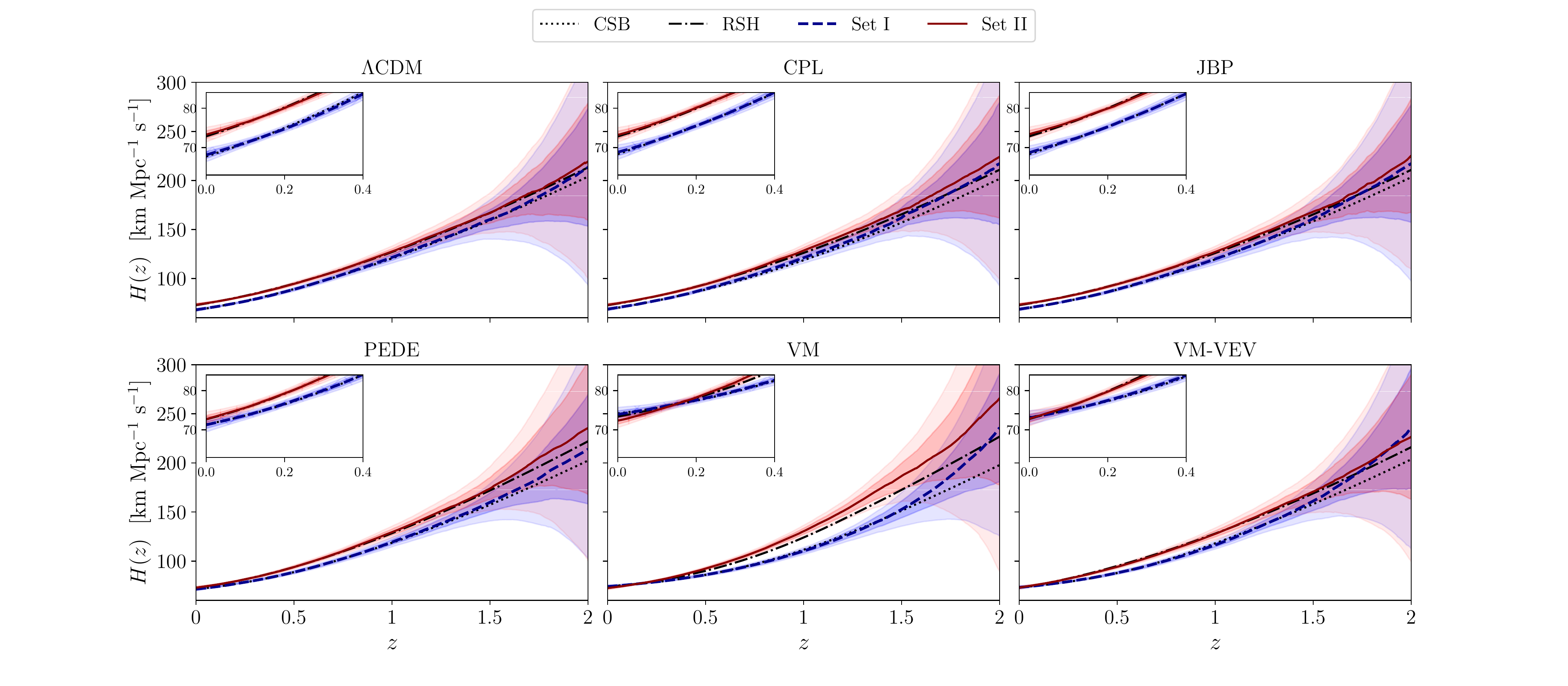}
\caption{Plots for the reconstructed $H(z)$ vs redshift $z$ from $\sim1000$ GW events detectable by ET for a $\sim3$-year mission duration in the redshift range $0<z<2$. Here `CSB' \& `RSH' denotes the fiducial Hubble function obtained using respective dataset combinations. Set-I \& Set-II represent the reconstructed functions with ET mock catalogs generated from the 'CSB' and 'RSH' constraints as fiducials.\label{fig:Hz_ET_plot}}
\end{figure}

\begin{figure}[!t]
\includegraphics[width=\textwidth]{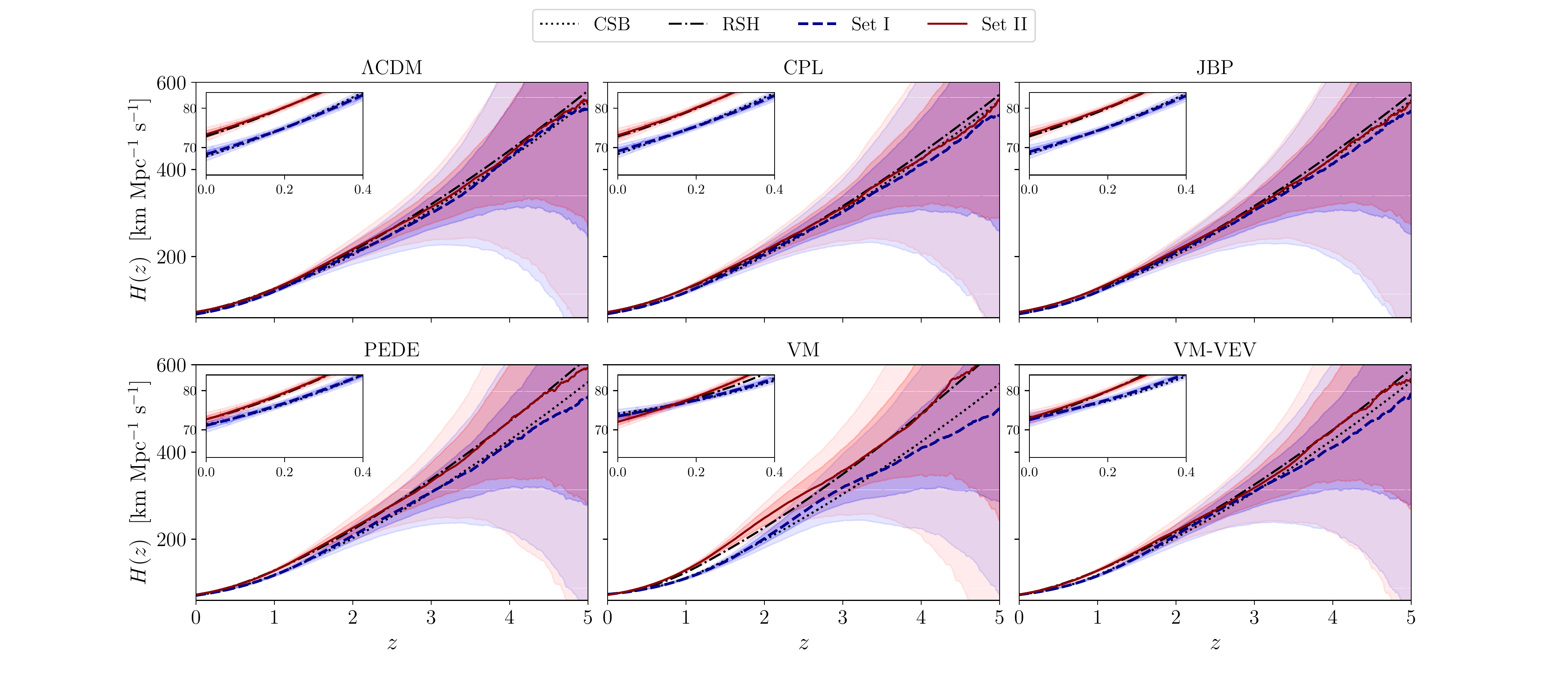}
\caption{Plots for the reconstructed $H(z)$ vs redshift $z$ using $\sim80$ GW events from the ``No Delay'' source population detectable by eLISA for a $\sim10$-year mission duration in the redshift range $0<z<5$.\label{fig:Hz_eLISA_plot}}
\end{figure}

\begin{figure}[!t]
\includegraphics[width=\textwidth]{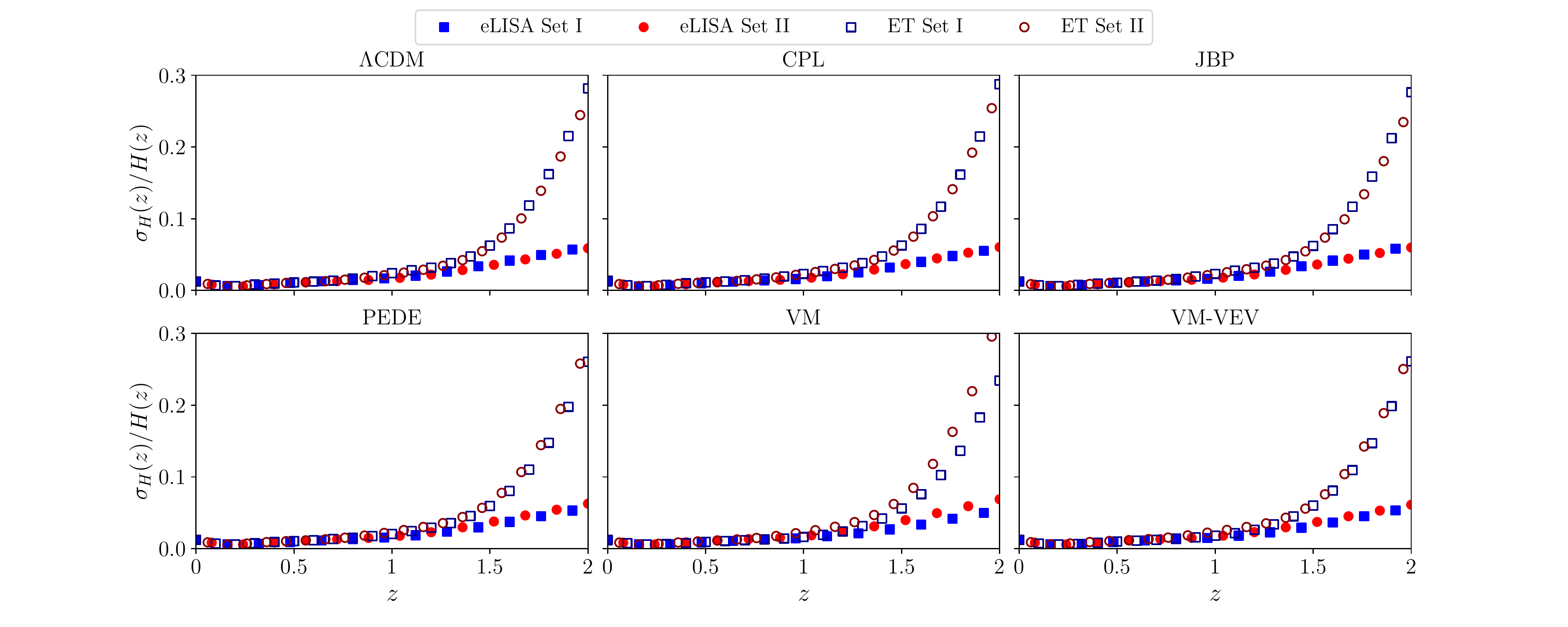}
\caption{Comparison of the errors in the reconstruction of $H(z)$ between ET and eLISA in the redshift range $0<z<2$ for mission particulars as outlined in Figs. \ref{fig:Hz_ET_plot} and \ref{fig:Hz_eLISA_plot}.}\label{fig:sigHz_compare}
\end{figure}

Finally, to put down the results obtained by this reconstruction exercise in a convenient way, we plot the reconstructed Hubble parameter $H(z)$ as a function of redshift. Fig. \ref{fig:Hz_ET_plot} shows the evolution of the Hubble parameter over a redshift range of $0<z<2$ detectable by ET for a $\sim3$-year mission duration for each of the six cosmological models under study. Fig. \ref{fig:Hz_eLISA_plot} shows the reconstructed $H(z)$ from simulated events detectable by eLISA with the ``No Delay'' MBHB source population for a $\sim10$-year mission duration, in the redshift range $0<z<5$. Each of the $H(z)$ plots have insets magnifying the reconstruction in the redshift range $0<z\lesssim 0.3$. For each plot, the dashed and solid lines correspond to the averaged reconstructed $H(z)$ curve from Set-I and Set-II mock catalogs. The shaded regions with $\times$ and $+$ hatches show the 1$\sigma$ and 2$\sigma$ confidence levels for Set-I and Set-II respectively. The fiducial Hubble function constrained using the CSB and RSH dataset combinations are plotted with dotted and dash-dot lines. We also plot the comparative constraining power on $H(z)$ between the two missions under consideration in Fig. \ref{fig:sigHz_compare}.

We further investigate the effects of mission duration, and hence the number of detected events, on the mission's potential in constraining $H_0$. We show how the reconstructed $H_0$ varies with the number of detected events by ET in Fig. \ref{fig:ET_whisker_plot}, and eLISA in Fig. \ref{fig:eLISA_whisker_plot}. Such a demonstration helps in adding perspective to whether and how these two upcoming future surveys will help in improving the constraints on $H_0$, in comparison to the latest constraints from currently available datasets. The square and circular markers with error bars represent the respective mean values and 1$\sigma$ uncertainties of the reconstructed $H_0$ from eLISA and ET, assuming both Set-I and Set-II mocks. We further indicate the early-time constraints from CSB datasets in blue color with $\times$ hatches, whereas the late-time ones using RSH data are denoted in red color with $+$ hatches respectively for the different models under consideration.

\section{Analysis and Discussion}\label{sec:analysis_discussions}

The present analysis reveals that GP can be a useful tool to reconstruct the Hubble parameter for future GW missions, at least to a moderately good redshift range. Our results also demonstrate the level of precision at which these next-generation missions should be able to constrain both the redshift evolution of the Hubble parameter $H(z)$ and its present value $H_0$.  

The $1\sigma$ error of the reconstructed value of $H_0$ shown in Fig. \ref{fig:ET_whisker_plot} and \ref{fig:eLISA_whisker_plot} generically decreases with an increase in the number of events, irrespective of the choice of the background cosmological model, the GW mission, and the set of fiducial parameter values used to generate the catalogs. It reveals the constraining power of these two missions for $H_0$ with relatively less error than current observations. Besides, the mean value of the reconstructed $H_0$ typically falls within its corresponding fiducial bounds in a stable manner for a sufficiently large number of events and beyond. These results indicate a faithful reconstruction. The saturation of the error with the number of events is, of course, set by the specifications of the individual mission. 

\subsection{Comparison between eLISA and ET}

\begin{figure}[!htb]
\gridline{\fig{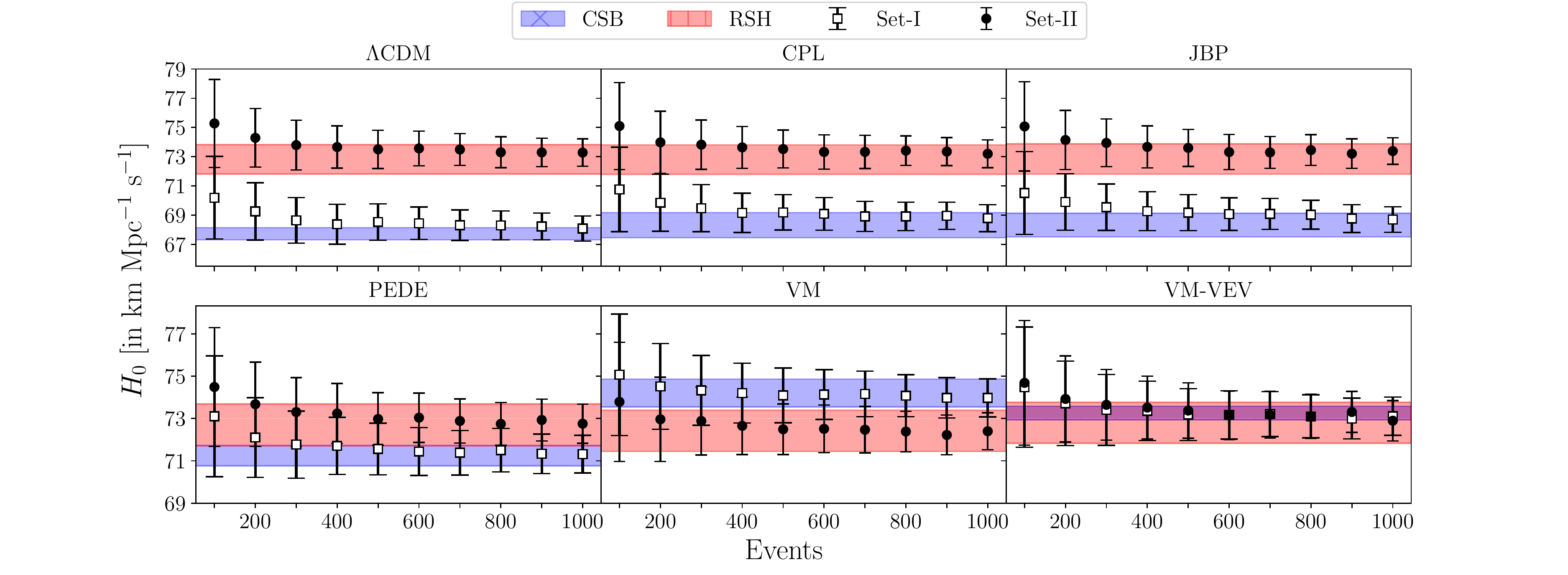}{\textwidth}{}}  \vspace{-0.75cm} 
\caption{Values of the reconstructed $H_0$ considering a variable number of GW events with EM counterparts detectable by ET.\label{fig:ET_whisker_plot}}
\end{figure}

\begin{figure}[!htb]
\gridline{\fig{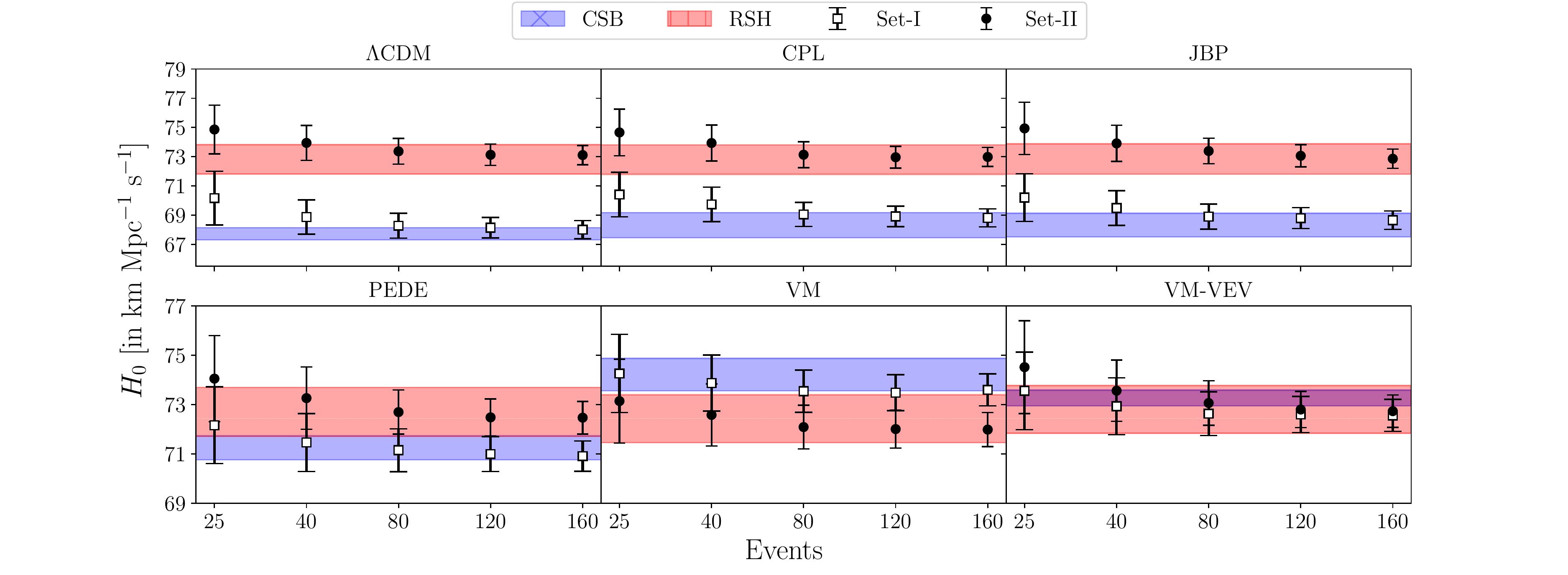}{\textwidth}{}}  \vspace{-0.75cm}
\caption{Values of the reconstructed $H_0$ considering variable mission duration for ``No Delay'' bright siren source type detectable by eLISA.\label{fig:eLISA_whisker_plot}}
\end{figure}

Coming to a comparison between these two particular GW missions, we first note that while ET is expected to detect a significantly higher number of bright sirens than eLISA up to $z\sim2$, a typical $\sim10$-year eLISA mission is predicted to constrain $H_0$ almost as precisely as ET, with longer mission durations generically tending to perform better than ET. This trend holds in a robust fashion for both sets of mock catalogs. The superior performance of eLISA can be attributed partly to its enhanced instrumental sensitivity, and partly to its ability of probing higher redshifts up to $z\sim8$ in search of standard sirens. The impact of the former factor is further visible in the plots of the reconstructed redshift evolution of $H(z)$ in the range $0<z<2$ for ET and eLISA (from Figs. \ref{fig:Hz_ET_plot} and \ref{fig:Hz_eLISA_plot}), where the reconstruction in case of eLISA is seen to be much tighter compared to the one in case of ET (Fig. \ref{fig:sigHz_compare}). Nonetheless, a direct comparison between the expected number of events, detectable by ET ($\sim1000$) and eLISA ($\sim25$), for a $\sim3$-year mission duration (from Figs. \ref{fig:ET_whisker_plot} and \ref{fig:eLISA_whisker_plot}) shows better performance of the former compared to the latter.

\subsection{Comparison among cosmological models}

\begin{figure}[!htb]
\gridline{\fig{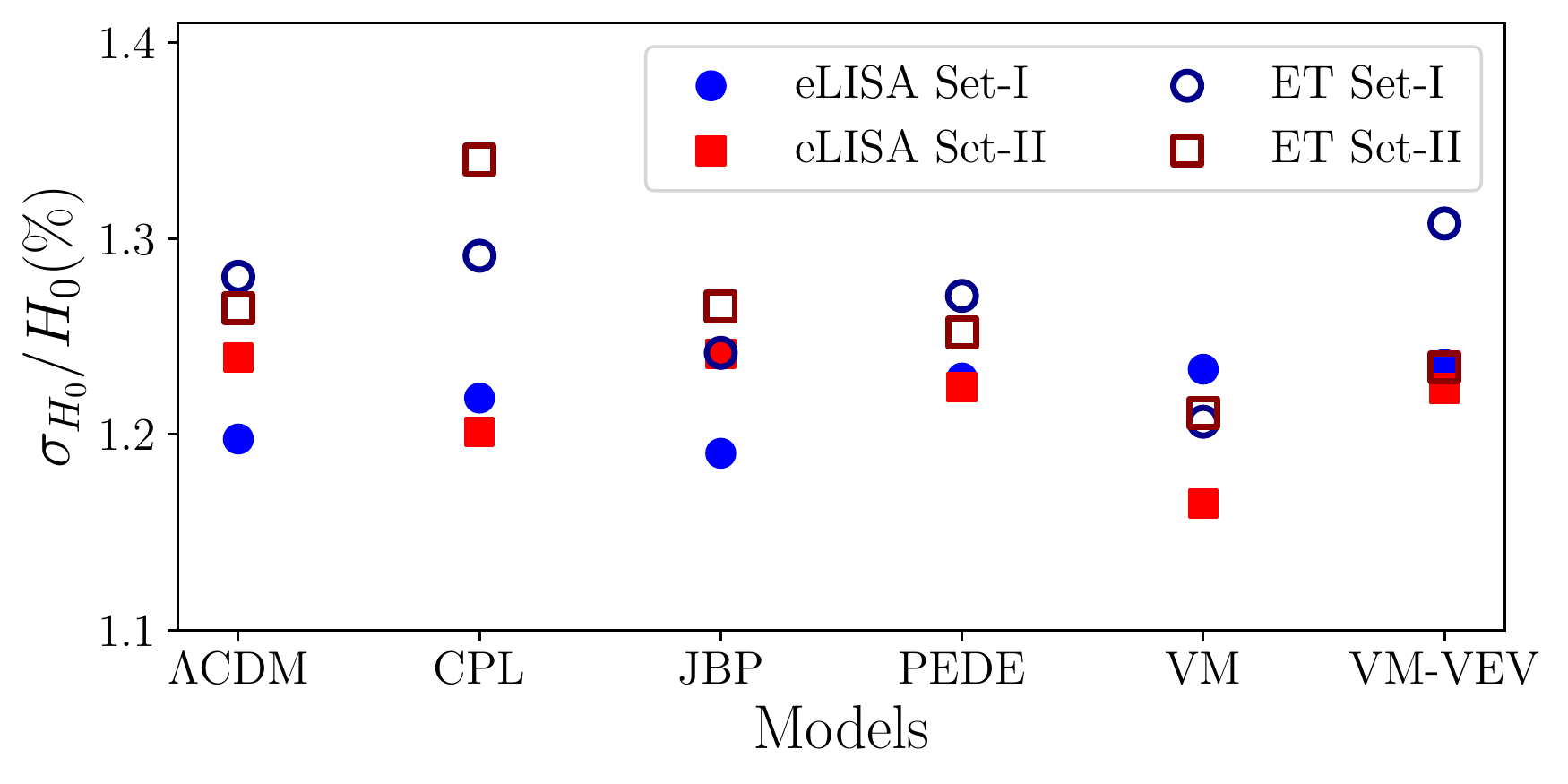}{0.45\textwidth}{}\hspace{-1cm}
          \fig{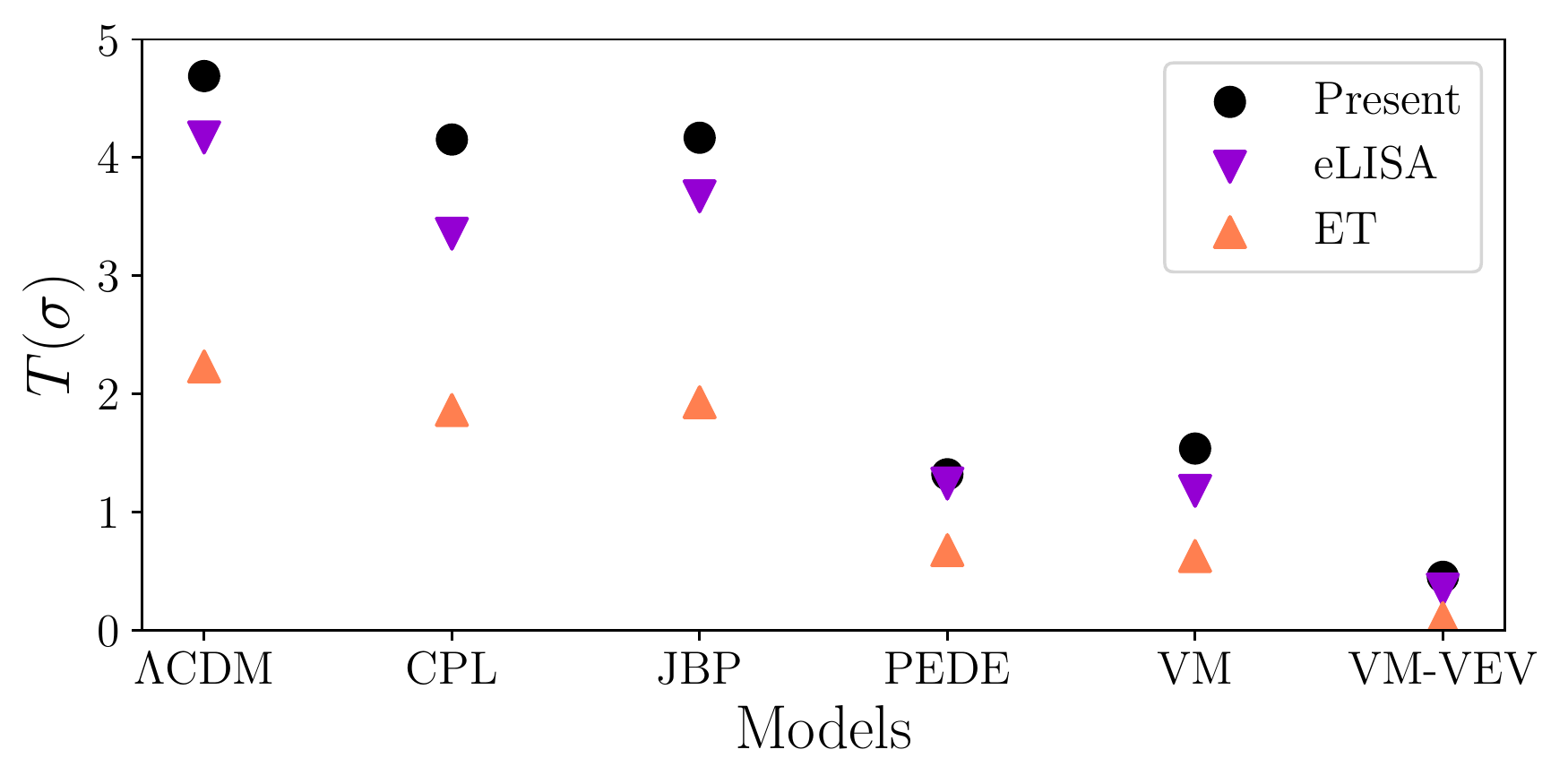}{0.45\textwidth}{}}\vspace{-0.85cm}
\caption{Plots showing a comparison of the relative uncertainty on reconstructed $H_0$ in the left panel. The right panel presents the resulting $H_0$ tension in light of eLISA and ET compared to the existing tension for different models.}\label{fig:forecast_plot}
\end{figure}

Based on the results obtained, we notice a few characteristic trends from Figs. \ref{fig:ET_whisker_plot} and \ref{fig:eLISA_whisker_plot}. For $\Lambda$CDM, CPL, and JBP, the reconstructed mean $H_0$ alongside its $1\sigma$ uncertainty tends towards a marginally higher value compared to the early-time prior in the case of ET, while the reconstruction corresponding to the late-time prior shows no significant shift. This is somewhat interesting from the perspective of the Hubble tension, although further studies with alternative ML algorithms are required in order to make any strong comments in that direction. The same mean-shifting tendency is visible to some extent in the case of eLISA as well for a realistic 10$-$15 year mission duration, albeit with tighter error bars which make the reconstructed values more consistent with the fiducials than in the case of ET. For the PEDE model, the opposite tendency is observed in the case of both ET and eLISA with an increasing number of events, as the reconstructed $H_0$ shifts marginally towards lower values for both early and late-time priors. This effect is not very significant though, as it still remains largely consistent with the fiducial which also includes the reconstructed mean. The VM model displays mean-shifting characteristics quite similar to PEDE, with the only difference being that the early-time fiducial value of $H_0$ in the case of VM is higher than the late-time one, as given by the MCMC results. Finally, for the VM-VEV model, the reconstructed mean value of $H_0$ falls almost outside the 1$\sigma$ bounds of the corresponding early-time fiducial but remains consistent with the latter within $1\sigma$. However, as the early-time prior itself falls entirely within the bounds of the late-time one with which both of the reconstructed $H_0$ values are entirely consistent, there is no statistically significant surprise in the outcome. We summarize the status of the $H_0$ tension for the different models with respect to present and mock GW datasets in the right panel of Fig. \ref{fig:forecast_plot}, where $T=\Delta\nobreak{}H_0/{\sigma}_{_{H_0}}$ is the 1-dimensional Gaussian tension metric.

\subsection{Comparison between fiducials}

For a given background model, the two separate reconstructed trajectories of $H(z)$ are unsurprisingly distinct and clearly dependent on the choice of the fiducial values used to generate the two different sets of catalogs. However, the reconstructed uncertainties in $H_0$ do not depend on or distinguish between the choice of the late-time or early-time fiducials for any given model, as shown in the left panel of Fig. \ref{fig:forecast_plot}. Besides, under the same choice of fiducials, the reconstructions obtained for the two different missions are mutually consistent for each of the models considered.  This provides another reality check on the fidelity of the GP reconstruction technique.  

Based on the reconstructed $H(z)$ plots in Figs. \ref{fig:Hz_ET_plot} and \ref{fig:Hz_eLISA_plot}, a few more model-specific features can be summarized. A unique feature of the VM model appears in the form of a crossing between the two mean trajectories of $H(z)$ for the early and late-time fiducials. This occurs in the approximate redshift region $0.1<z<0.2$, beyond which the reconstruction based on the late-time prior gives a higher value of $H(z)$ compared to the early-time one. While the VM model is capable of resolving the $H_0$ tension at present within $1\sigma$, the reconstructed history thus obtained indicates a tension in $H(z)$ at higher redshifts. This tension persists up to $z\sim1.5$ in the case of ET and $z\sim2$ in the case of eLISA, beyond which the evolutionary histories of $H(z)$ are once again consistent with each other within $1\sigma$ but solely on the basis of widened error bars (the mean curves continue to be significantly apart). None of the other models shows such a pronounced discrepancy at higher redshifts. While the mean $H(z)$ curves for PEDE and VM-VEV are also somewhat distant from each other, the difference is not statistically significant due to the large uncertainties involved. For $\Lambda$CDM, CPL, and JBP, the two reconstructed trajectories are comparatively more concordant with each other, in spite of the worse status of these models as far as the present $H_0$ tension is concerned. 


\section{Conclusions}\label{sec:conclusion}

In this article, we analyze the robustness of Gaussian processes (GP) in reconstructing the Hubble parameter in light of future gravitational wave (GW) missions. We employ the following methodology, which enables us to assess the efficacy of GP in a more or less robust and unbiased manner:
\begin{itemize}
    \item We focus on the ground-based ET and the space-based eLISA as two representatives of next-generation GW detectors. These missions aim to detect gravitational waves originating from different astrophysical phenomena by probing different redshift ranges. This helps mitigate mission-specific dependencies within our general conclusions which are based on a comparative study between the individual reconstruction results for both missions. 
    \item We consider six different background cosmologies (including the concordance $\Lambda$CDM model as the benchmark). The chosen scenarios incorporate late-time modifications on top of the vanilla 6-parameter picture that subsequently leads to considerable variations in the theoretical $d_L$ vs $z$ behavior. This, in turn, allows us to study various representative cases of the same and arrive at conclusions which hold generically across a variety of cosmological models (Table \ref{tab:models}).
    \item For each mission and model under consideration, we generate two sets of mock catalogs (Set-I and Set-II) using the constraints obtained from currently available early-time (CSB) versus late-time (RSH) datasets as fiducials (Table \ref{tab:study_cases}). 
    \item We examine their relative effect on the reconstructed results, owing to the tension between early and late-time constraints that persists particularly in the value of the Hubble constant. We have resorted to this disparate analysis involving two distinct fiducials, assuming neither set to be indisputably correct in an \textit{a priori} fashion.
\end{itemize}

The reconstructed plots of the Hubble parameter $H(z)$ are strongly dependent on the choice of the fiducials used to generate the mock catalogs. However, under the same choice of fiducials, the reconstructions corresponding to the two different missions are consistent with each other for each of the models considered, which indicates a faithful reconstruction process. Upon varying the number of detectable events or (equivalently) the mission duration, the presently observed mean value of the reconstructed $H(z)$ and its associated $1\sigma$ error also show generic trends for both missions in accordance with their specifications. In particular, our results confirm the following:
\begin{itemize}
\item GP is a powerful and efficient non-parametric tool which can be used to reconstruct the Hubble diagram based on standard siren datasets from multiple next-generation GW detectors, and hence constrain the value of the Hubble constant based on GW luminosity distance data at intermediate redshifts. 
\item Concerning precision, the results obtained from a $\sim10$-year eLISA mission are likely to be at par with those from a $\sim3$-year ET run. Thus, these future missions have the potential to constrain cosmological parameters and provide bounds that would be competitive to present-day constraints.
\item If the missions run for a longer duration, the reconstruction results show further improvement. For example, the relatively tighter constraint on $H_0$ from a $\sim15$-year eLISA mission can be readily compared with those from shorter runs in Fig. \ref{fig:eLISA_whisker_plot}.
\end{itemize}

However, in this work, we have chosen not to combine the simulated data from eLISA and ET because of two primary sources of uncertainty: the dissimilar systematics of the missions, and the questionable physical validity of combining luminosity distance data obtained with distinct instruments. We plan to explore this avenue in future work. 

Moreover, a natural question that arises here is: What about other future GW missions, such as the DECi-hertz Interferometer Gravitational-wave Observatory (DECIGO) \citep{DECIGO1,DECIGO2,DECIGO3}, the Big Bang Observer (BBO) \citep{bbo1,bbo2}, TianQin \citep{TianQin:2015yph}, Cosmic Explorer \citep{CosmicExplorer:2019iox}, Taiji \citep{Taiji:2018tsw} as well as ongoing GW missions such as the LIGO-Virgo-KAGRA \citep{LVK:2013rdx} network and the Pulsar Timing Array \citep{PTA:2009yy}? We have already carried out some initial checks with DECIGO, which proposes a large number of detections in the redshift range $0<z<5$ with a very high signal-to-noise ratio (SNR). We found that the reconstructed $H(z)$, and hence $H_0$, obtained from a simulated catalog strongly mimic the fiducial values, albeit with increased precision. This shows the reconstruction efficacy of GPs. At the same time, averaging over other realizations of the catalogs for DECIGO is not expected to give much variability. However, given the success of GP for eLISA and ET, it would be intriguing to find out the prospects of other GW missions that plan to probe events at relatively higher redshifts. 

Furthermore, in the present study, we have exclusively focused on bright standard sirens, i.e. those with detectable electromagnetic counterparts. This choice is reasonable because upcoming optical and radio observatories like LSST, SKA+ELT, and THESEUS/ATHENA are expected to identify transient EM emissions accompanying gravitational wave (GW) events. This would aid in pinpointing the source galaxies' locations and redshifts even at high redshifts. Recent reports from the operational James Webb Space Telescope (JWST) show promise in probing luminous sources beyond $z\sim10$ via EM observations, determining their redshifts through spectroscopy or photometry \citep{jwst1, jwst2, jwst3, jwst4, jwst5}. Future observatories' improved capabilities are anticipated to enhance collaboration between GW and EM facilities, enabling real-time study of high-redshift multi-messenger events and subsequent follow-ups of host galaxies. If, on the other hand, dark sirens could also be localized with sufficient precision using galaxy cross-correlation or any other viable technique(s) \citep{galaxycorr1,galaxycorr2,galaxycorr3}, their addition would significantly increase the number of detectable events at higher redshifts, thus bolstering the fidelity of the results obtained through our adopted methodology. Such prospects have attracted considerable attention in recent years, in light of a number of proposed ground-based and space-based GW missions \citep{Seymour:2022teq,lisa_taiji,Jin:2023sfc,et_dark_1,et_dark_2}. As far as the reconstruction methodology demonstrated in this work is concerned, these are a couple of other challenging directions that would be interesting to pursue in near future.

Finally, the applicability of other machine learning techniques besides GP to the reconstruction of $H(z)$, based on standard siren datasets either from eLISA/ET or from some of the other future GW missions mentioned above, is another potentially interesting field to shed light on.
\\\\

\begin{acknowledgments}
We thank the anonymous reviewer for their valuable suggestions towards the improvement of the manuscript. PM thanks ISI Kolkata for financial support through Research Associateship. RS thanks ISI Kolkata for financial support through Senior Research Fellowship. AB thanks CSIR for financial support through Junior Research Fellowship (File no. 09/0093(13641)/2022-EMR-I). SP thanks the Department of Science and Technology, Govt. of India for partial support through Grant No. NMICPS/006/MD/2020-21. 
\end{acknowledgments}

%



\software{\href{https://github.com/lesgourg/class_public}{\textit{CLASS}} \citep{CLASS1,CLASS2},  
          \href{https://github.com/brinckmann/montepython_public}{\textit{MontePython}} \citep{MontePython1,MontePython2}, 
          \href{https://github.com/dfm/emcee}{\textit{emcee}} \citep{Foreman_Mackey_2013},
          \href{https://github.com/carlosandrepaes/GaPP}{\textit{GaPP}} \citep{Seikel:2012uu}
          }

\bibliography{references}{}
\bibliographystyle{aasjournal}



\end{document}